\documentstyle[a4,12pt,psfig]{article}
\newcommand{\al}{\alpha}
\newcommand{\bt}{\beta}

\newcommand{\be}{\begin{equation}}
\newcommand{\ee}{\end{equation}}
\newcommand{\ba}{\begin{eqnarray}}
\newcommand{\ea}{\end{eqnarray}}

\newcommand{\vsa}{\vspace{0.4cm}}
\newcommand{\vsb}{\vspace{0.8cm}}

\begin{document}
\begin{titlepage}
\begin{flushright}
hep-th/9601025 \\
THU-96/03\\
January 1996
\end{flushright}
\vsa
\begin{center}
{\large\bf Pauli-Lubanski scalar in the Polygon Approach to 2+1-Dimensional
Gravity
         \vsa\vsb\\}
           M. Welling\footnote{E-mail: welling@fys.ruu.nl} and M. Bijlsma
           \vsa\vsb\\
   {\it Institute for Theoretical Physics\\
     University of Utrecht\\
     Princetonplein 5\\
     P.O.\ Box 80006\\
     3508 TA Utrecht\\
     The Netherlands}\vsb\vsa\\
\end{center}
\begin{abstract}
In this paper we derive an expression for the conserved Pauli-Lubanski scalar
in 't Hooft's polygon approach to 2+1-dimensional gravity coupled to point
particles. We find that it is represented by an extra spatial shift $\Delta$ in
addition to the usual identification rule (being a rotation over the cut). For
two particles this invariant is expressed in terms of 't Hooft's phase-space
variables and we check its classical limit.
\end{abstract}

\end{titlepage}
\section{Introduction}
\begin{figure}[t]
\centerline{\psfig{figure=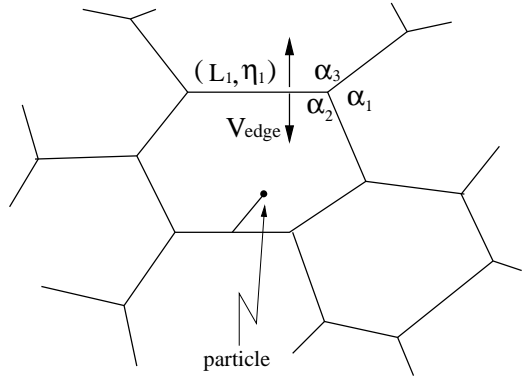,angle=-90,height=5cm}}
\caption{Tesselation of space by polygons.}
\label{tesselation}
\end{figure}

In 1992 't Hooft introduced the polygon approach to 2+1-D gravity
\cite{tHooft}. By using the fact that gravity in 2+1 dimensions has only matter
degrees of freedom he managed to write down an exact solution to 2+1-D gravity
coupled to point particles. In this formulation one tesselates a time-slice
with polygons (see figure \ref{tesselation}) and defines a Lorentzframe on each
polygon. The velocity of the boundary between two adjacent polygons (generally
with different Lorentzframes) is determined by the equation $t_1=t_2$ \,\, i.e.
the time on both polygons must be equal. This ensures that no time jumps take
place and the surface is a proper Cauchy surface. One can show that in this
case the boundary can only move perpendicular to itself with a constant
velocity. The lengths of the boundaries will be taken as the configuration
variables of phase space, and are denoted by $L_i$.

As a momentum variable conjugate to $L_i$, we must take $2\eta_i$, where
$\eta_i$ is the rapidity of the edge $L_i$ ($v_{\bf edge}=\tanh\eta$). Thus we
have: $\{L_i,2\eta_j\}=\delta_{ij}$. The line segments $L_i$ end either in a
vertex, where three edges meet (three-vertex), or in a particle (one-vertex),
as depicted in figure \ref{tesselation}. At the three-vertex the three
dimensional curvature must vanish as there is no matter present at this
location. Exploiting this fact one can derive a set of equations which can be
used e.g. to express the three angles $\alpha_i$ in terms of the adjacent
rapidities $\eta_j$. If there is a particle at the end of a line $L_i$ we pick
up
a nontrivial holonomy if we move around the particle. The length of an edge
$L_i$ may shrink or grow, and this can be described by a Hamiltonian
formulation. If we take the Hamiltonian as the sum of the angle deficits at the
particle-sites and at the three-vertices we generate the correct dynamical
equations:
\ba
\frac{d}{dt}L_i&=&\{H,L_i\}\\
\frac{d}{dt}\eta_i&=&\{H,\eta_i\}~~~~~~~~~(=0)
\ea
As the system evolves, one of the lengths $L_i$ may shrink to zero and a
transition to a new set of polygons takes place. The possible transitions are
listed in
\cite{tHooft}. Another aspect of the Hamiltonian formulation is the fact that
there are constraints on the variables. The constraints among the variables
result from the fact that the polygons must close. These constraints are first
class and generate time translations and Lorentz transformations of the
polygon. We also refer to the literature
for a detailed description of these constraints \cite{tHooft}. In this paper we
will actually concentrate only on one (open) polygon (figure \ref{Delta} ).

 The polygon formulation of 2+1-D gravity  was recently revisited by Franzosi
and Guadagnini who stressed the importance of the braid group \cite{Franzosi}.
Using holonomy loops they obtained a formula expressing the conservation of
energy-momentum
during particle interactions. In this paper we will derive a similar kind
of conservation law for the Pauli-Lubanski scalar ${\cal J}$. It is long known
that this conservation law exists \cite{Carlip}. In this paper we will
investigate this conservation law in the polygon approach where we cannot allow
for time jumps. We find that it is represented by an extra spatial shift
$\Delta$ in the identification over the cut.
\section{Observables in 2+1-Dimensional Gravity}
The usual way to describe point particles in 2+1-D gravity is by cutting out
wedges from space-time and identifying the edges according to certain rules.
For instance in the case of two moving particles we have the following
identification rule \cite{beginarticle}:
\ba
\tilde{x}&=&Lx+q\\
         &=&(B_1R_1B_1^{-1}B_2R_2B_2^{-1})x\nonumber\\
&+&(-B_1R_1B_1^{-1}B_2R_2B_2^{-1}a_2+B_1R_1B_1^{-1}(a_2-a_1)+a_1)\nonumber
\ea
Here $a_i$ are the locations of the particles, $B_i$ are boost matrices in
SO(2,1), $R_i$ are rotation matrices over
an angle $m_i$ (equal to the mass of particle $i$), $L$ is a Lorenz
transformation of the form: $L=BRB^{-1}$ and $q$ is a translation vector. We
have chosen units such that $8\pi G=1$. In the above formulas we omitted the
indices. In \cite{beginarticle} it was found that the angular momentum is given
by $q^0$, which is a quantity that transforms under Lorentz transformations and
translations. The "observables" in this theory must be given by the invariants
of the Poincar\'e group. There are two such invariants \cite{Carlip}, namely:
\begin{enumerate}
\item \be{\bf Tr}L=1+2\cos M\ee
where $M$ is the total mass of the system, and
\item \be {\cal J}=\frac{-1}{2\sin M}\varepsilon^{abc}L_{ab}q_c=[B^{-1}_{\bf
com}q]^0\ee
\end{enumerate}
The matrix $B_{\bf com}$ is the boost-matrix for the effective center of mass
particle. The second expression can be derived using $L=B_{\bf com}R(M)_{\bf
com}B^{-1}_{\bf com}$ and $LJ^aL^{-1}=J^bL_b^{~~a}$ with
$(J^a)^b_{~~c}=(\varepsilon^a)^b_{~~c}$. From ii) we see that the second
invariant is really proportional to the angular momentum in the {\em center of
mass frame}. It is in fact the Pauli-Lubanski scalar, which is proportional to
the spin of a particle in its rest frame.

\section{Pauli-Lubanski Scalar in the Polygon Approach}
In this section we will see how ${\cal J}$ is represented in the polygon
approach to 2+1-D gravity. As mentioned in the introduction the polygon
approach is a Cauchy formulation and no closed timelike curves are allowed by
construction. It also implies that we must choose the cuts in such a way that
there are no time jumps anywhere in the plane.

\begin{figure}[b]
\centerline{\psfig{figure=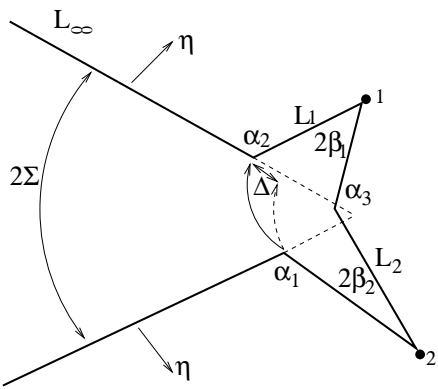,angle=-90,height=5cm}}
\caption{Two particles with angular momentum.}
\label{Delta}
\end{figure}

The total energy of the system is given by the total angle deficit of an
"effective" center of mass particle and can be expressed in terms of the rest
masses of the particles and the momenta $\eta_i$ across the edges. Similarly we
expect that ${\cal J}$ will be given by the spin of the center of mass particle
and that it can be expressed in terms of the $L_i$ and $\eta_i$ of the
constituent particles. Although we know that close to the (spinning) center of
mass particle (where closed timelike curves are possible)
the Cauchy construction of 't Hooft is not possible, we still expect that far
away from the particles the identification rule for a single center of mass
particle is valid as an effective description for the system of particles.
The general identification rule
for a moving and spinning particle situated at the origin is:
\be
\tilde{x}=BRB^{-1}x+Bs~~~~~~~~~~~s=(S,0,0)  \label{trafo}
\ee
We want to choose the cut in such a way that (at $t=0$) there is no time jump
across this cut:
\be
\tilde{x}^0=[BRB^{-1}x+Bs]^0=0\label{con}
\ee
Next we choose this particle to move in the positive $x$-direction. The
condition (\ref{con}) then gives the following line of points (parametrized by
$\ell$) that will not experience a time jump under the transformation
(\ref{trafo}):
\ba
x&=&\ell\\
y&=&\ell\tan\Sigma~-\frac{S}{v\sin M}\nonumber
\ea
with $\tan\Sigma=\cosh\xi\tan\frac{M}{2}$ and $v=\tanh\xi$. This line is mapped
by (\ref{trafo}) to:
\ba
\tilde{x}&=&\ell-\frac{S}{\sinh\xi}\\
\tilde{y}&=&-\ell\tan\Sigma~-\frac{\cot M~S}{v}\nonumber
\ea
This mapping is pictured in figure \ref{map}. First we note that the
point of intersection of the two lines in figure \ref{map} is not at the
position of the particle itself. Secondly, the mapping is not only a rotation
over the angle $2\Sigma$ but also contains a shift over a distance
$\Delta=S/(\sinh\xi\cos\Sigma)$. We see that close to the particle the
construction becomes pathological as anticipated. This, however, presents no
problem as this part of space
will be replaced by the space of the moving particles without spin. Far away
from the particles however, ${\cal J}$ is still given by:

\begin{figure}[t]
\centerline{\psfig{figure=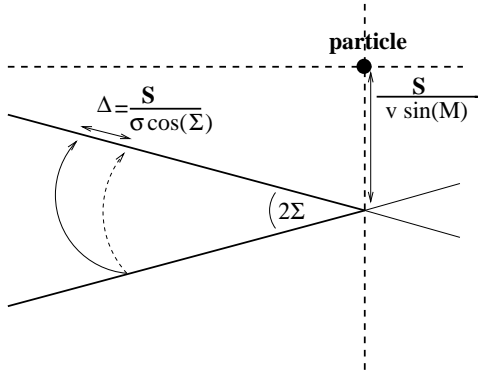,angle=-90,height=5cm}}
\caption{Wedge cut out of space-time for a moving spinning particle
($\sigma=\sinh\xi$).}
\label{map}
\end{figure}

\be
 {\cal J}=\Delta\cos\Sigma\sinh\xi
\ee
We can, in the case of two particles, express this quantity ${\cal J}$ in terms
of the polygon variables $L_i$ and $\eta_i$ (figure \ref{Delta}) as follows;
\be
{\cal J}=-2\sinh\xi\{L_1\sin\bt_1\sin(\bt_2+\frac{1}{2}(\al_2-\al_1-\al_3))
-L_2\sin\bt_2\sin(\bt_1+\frac{1}{2}(\al_1-\al_2-\al_3))\}
\ee
The angles $\al_i$ and $\bt_i$ can be expressed in terms of the rapidities
defined across the edges (see (\cite{tHooft})). Furthermore,
$\sinh\xi=\sinh\eta/ \sin\frac{1}{2}M$, where $\eta$ is the rapidity over the
edge $L_\infty$ and $M$ is the total mass of the system, which can also
be expressed in terms of the rest masses and rapidities of the constituent
particles.
\newpage
In order to check the classical limit we have to take:
\ba
\sinh\xi&\rightarrow & v_{\bf com}\\
2\sin\bt_i&\rightarrow & m_i
\ea
and find (figure \ref{klass}):
\begin{figure}[t]
\centerline{\psfig{figure=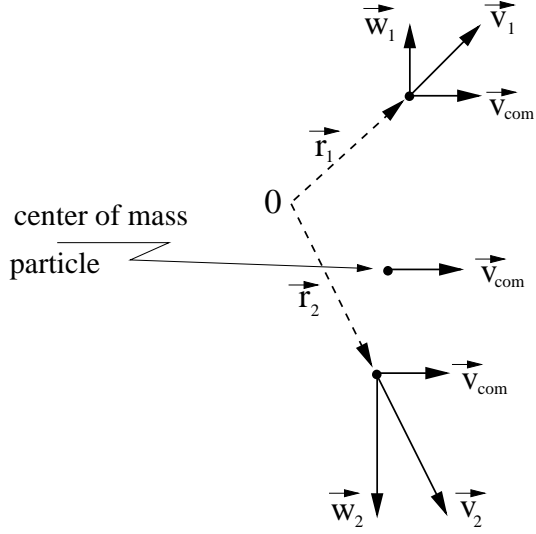,angle=-90,height=7cm}}
\caption{Classical limit of figure \ref{Delta}.}
\label{klass}
\end{figure}
\be
{\cal J}=-\sum_{i=1}^{2}m_i\vec{r_i}\times\vec{v}_{\bf com}
\ee
If we choose the origin as in figure \ref{klass}, such that the angular
momentum vanishes in this (boosted) frame, this quantity indeed represents the
total angular momentum of the two particles in their center of mass frame,
written in terms of the boosted variables:
\be
{\cal J}=\sum_i m_i \vec{r}_i\times\vec{w}_i =\sum_i m_i
\vec{r}_i\times(\vec{v}_i-\vec{v}_{\bf com})= -\sum_i m_i\vec{r}_i\times
\vec{v}_{\bf com}
\ee
where $\vec{w}_i$ is the velocity in the c.o.m. frame and $\vec{v}_i$ is the
velocity in the boosted frame.

\section{Discussion}
The first step towards quantizing 2+1-D gravity in the polygon approach has
been made recently by 't Hooft \cite{dis}. In this paper the quantization of
one particle in its own gravitational field is considered. Surprisingly, he
finds that space-time is discretized in this case. The next logical step would
be to quantize the two particle problem. As it is very convenient to have a
Hamiltonian formulation as a starting point for quantization, the polygon
approach seems to be the natural choice for describing the system. The two
particle system is expected to have two observables: energy and angular
momentum of the center of mass frame (or Pauli-Lubanski scalar). These
conserved quantities must be "measurable" at spatial infinity. In the polygon
approach energy was identified with the total angle deficit of space but no
geometrical quantity was identified as the Pauli-Lubanski scalar. We found in
this paper that it is given by a spatial shift in the identification rule at
spatial infinity.

\end{document}